\documentclass[conference]{IEEEtran}
\IEEEoverridecommandlockouts
\usepackage{cite}
\usepackage{amsmath,amssymb,amsfonts}
\usepackage{algorithmic}
\usepackage{graphicx}
\usepackage{subcaption}
\usepackage{textcomp}
\usepackage{xcolor}
\usepackage{siunitx}
\usepackage{soul}
\usepackage{threeparttable} % To get footnotes in tables
\usepackage[numbers]{natbib}

\usepackage{color}
\usepackage{xcolor}
\newcommand{\minisection}[1]{\vspace{0.05in} \noindent {\bf #1} \ }

\usepackage{todonotes}

\def\BibTeX{{\rm B\kern-.05em{\sc i\kern-.025em b}\kern-.08em
    T\kern-.1667em\lower.7ex\hbox{E}\kern-.125emX}}
\begin{document}

\title{Towards Informed Design and Validation Assistance in Computer Games Using Imitation Learning}
%\title{Imitation Learning for Game Design and Gameplay Validation: User study and experiments}
% \title{Towards Imitation Learning for Video Game Design and Validation}
% \title{Challenges and Opportunities for Using Imitation Learning for Video Game Design and Validation}
% \title{Challenges and Opportunities for Creating Imitative Agents for Game Design and Validation}
% \title{Recent Research and Future Directions in Imitation Learning for Video Game Design and Validation}
% \title{Towards Informed Design and Validation Assistance in Modern Games Using Imitation Learning}
% \title{Recent Research and Future Directions in Design and Validation Assistance in Modern Games using Imitation Learning}

\author{
Alessandro Sestini$^{1,2}$,
Joakim Bergdahl$^1$,
Konrad Tollmar$^1$,
Andrew D. Bagdanov$^2$,
Linus Gisslén$^1$
\\
\textit{$^1$SEED - Electronic Arts (EA)}\\
\textit{$^2$Università degli Studi di Firenze}
\\
\{jbergdahl, ktollmar, lgisslen\}@ea.com \\
\{alessandro.sestini, andrew.bagdanov\}@unifi.it
}

\maketitle

\begin{abstract}
% Alt1.
% In games, and many other domains, design validation and testing is a huge challenge as systems are growing in size and becoming infeasible to manually test. Real-time feedback would hugely benefit the production cycle and the quality of the product. In this paper we propose an automated approach for game design and gameplay validation. The method is data driven and purely based on demonstration and therefore should be easily accessible to non-ML experts. We deploy an imitation learning method to assess the validity of the approach. A survey based on answers from industry experts show promising results, and that this approach can be used to give feedback and help with analysis of game play. We also, with the help of the survey, give directions on which directions future research should take in order to maximize this data driven approach utility.

%% Alternative abstract
% Alt2.
In games, as in and many other domains, design validation and testing is a huge challenge as systems are growing in size and manual testing is becoming infeasible. This paper proposes a new approach to automated game validation and testing. Our method leverages a data-driven imitation learning technique, which requires little effort and time and no knowledge of machine learning or programming, that designers can use to efficiently train game testing agents. We investigate the validity of our approach through a user study with industry experts. The survey results show that our method is indeed a valid approach to game validation and that data-driven programming would be a useful aid to reducing effort and increasing quality of modern playtesting. The survey also highlights several open challenges. With the help of the most recent literature, we analyze the identified challenges and propose future research directions suitable for supporting and maximizing the utility of our approach.
\end{abstract}

\begin{IEEEkeywords}
Automated playtesting, imitation learning, game design
\end{IEEEkeywords}

\section{Introduction}
\label{sec:introduction}
%Problem statement
Modern games are often enormous and are growing exponentially in size, complexity, and asset count on every iteration. To ensure the game works as intended each time a level or level area is created or modified, there is a need for gameplay validation. Playtesting is one common procedure for assessing the quality of games. The testers check whether the game is completed, if it is fun and sufficiently challenging, or if it has problems like bugs or glitches. The process of game validation and testing is usually done either by an in-house quality verification (QV) team or by internal and external designated human testers. However, manual playtesting does not scale well with the size of the game and turnaround time -- the time between game design and feedback -- is often long.

\begin{figure}
    \begin{center}
    \includegraphics[width=0.85\columnwidth]{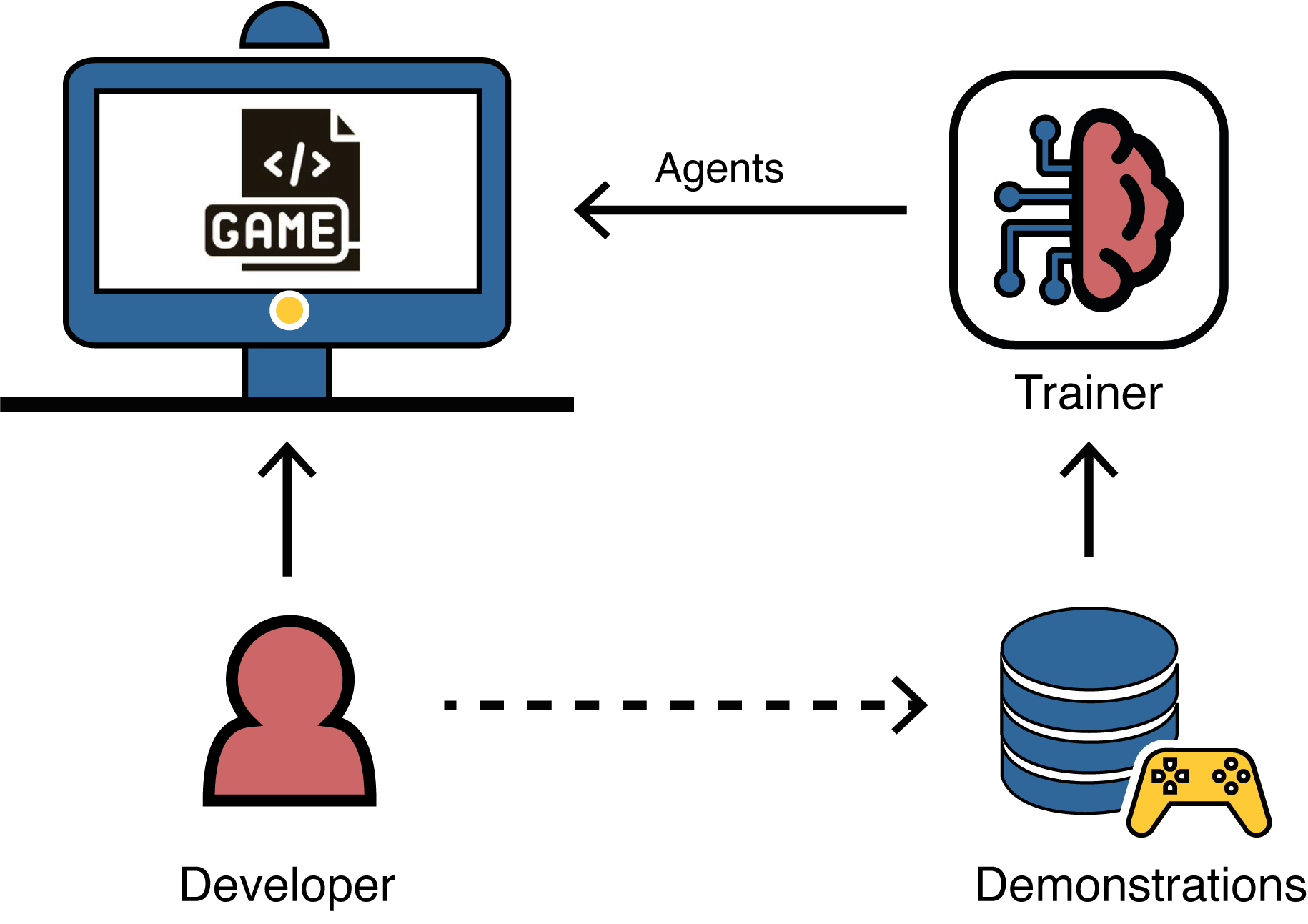}
    \end{center}
    \caption{ 
    % The process of game validation is currently made by external entities such as human playtesters and quality verification. 
    Our proposed data-driven approach lets designers perform automated game validation directly during design phase. The training is interactive as the algorithm allows designers to smoothly switch between designing the level, providing gameplay demonstrations and getting feedback from the trained testing agents.}
    \label{fig:teasing}
\end{figure}

%An argument for automated validation for game design
To be effective, game validation should be performed as early as possible and ideally by those who actually create the content that requires testing: the game creators. Note: game validation in this context is the process of evaluating that the game plays and ``feels'' like intended. For manual game testing, this is often limiting as bringing in human playtesters would be slow, expensive and inefficient. With automated game validation, it would be possible to bring in game playing agents directly into the development phase.

%Current reasearch: automated testing and validation
Recently, automated playtesting and validation techniques have been proposed to mitigate the need for manual validation in large games. Automated gameplay testing offers a fast and relatively cheap solution for testing games at scale. This is often done by crafting model-based bots and using them to perform automated playtesting \cite{pathos}. Another proposed solution as seen in recent research that also highlights some industry use-cases is to train self-learning agents with reinforcement learning \cite{improving, ccpt, ea_adversarial}. However, both of these approaches suffer from a few drawbacks. 
%Solution: Model based drawbacks
Model-based agents, that typically refers to bots with hand-crafted behaviors (i.e. achieved by scripting), require a certain level of domain knowledge and programming skills that game designers do not necessarily have. Moreover, the lack of generalization in this method might render the agents unusable if changes are made in the environment.
% Solution: RL drawbacks
Reinforcement learning, on the other hand, can learn to play the game without scripting, and even can be retrained when the environment change. However, reinforcement learning is sample-inefficient and arguably requires a high level of expertise in machine learning to effectively be used, e.g. for properly creating a well performing  reward function. Furthermore, reinforcement learning in general provides a low level of controllability as the agents will try to exploit the environment regardless of the intentions of the designer~\citep{faultyreward}. Additionally, making a game compliant with a reinforcement learning training setup is a considerable engineering effort potentially requiring intrusive changes to the game's source code.

%Prerequisite for a use-case
Ideally for an effective design tool for game designers, it should satisfy a set of identified requirements: 
\begin{itemize}
    \item imitation - the agent behaviors should be learned from demonstration. In this way the designer can show, rather than program, desired behaviors;
    \item efficiency - as a tool for real-time game validation, it needs to be close to instantaneous, therefore training time and required samples should be minimal;
    \item generalization - the method should be able to adapt to reasonable small design changes without retraining;
    \item controllability - having full control over behaviors can sometimes be crucial, especially when validating short sequences in games; and
    \item personas - the method should mimic different players' play-styles in order to lead to more meaningful testing; and
    \item machine learning expertise - no machine learning knowledge should be required (as this is not a standard tool in game design and therefore not common knowledge).
            
\end{itemize}

%Proposed solution
In this paper, we propose a data-driven approach for creating agent behaviors for automated game validation visible in Figure~\ref{fig:teasing}. Unlike methods where agents are manually programmed/scripted, this method uses imitation learning, in particular the DAgger algorithm~\cite{dagger}, to clone the behavior of the user. With this, we can leverage the expertise of the game developers while at the same time not require additional knowledge in programming or machine learning. The feedback loop for the designers can therefore be instantaneous with a much more efficient creation process than having to wait days and sometimes weeks for manual playtests. This study is made to either validate or refute above claims. The research questions we are focusing on are: can imitation learning be used as an effective game design tool? What improvements and research are needed to maximize its value as a tool? To explore this we showcase the possible use-cases for which level designers could use our data-driven approach. Further, we also propose a survey that explores how our method satisfies the requirements previously mentioned and gauges the interest of such a tool among professional developers. Based on the answers from the survey, we propose future research directions for improving game validation through imitation learning.

%Our hypothesis is that data-driven modelling, and in particular imitation learning (IL), can be used for game design testing and validation. 
%Conclusion and results of the paper
% The key contributions of this paper are:
% \begin{itemize}
%     \item an approach based on the DAgger algorithm \cite{dagger} to showcase the possible use-cases for which level designers could use this type of data-driven programming and why this approach satisfy all the requirements previously mentioned.
    
%     \item a survey that not only validates the approach, but also evaluates the interest in such a tool among professional developers.
    
%     \item a list of future research directions based on the answers we got from the survey to improve game validation through imitation learning. At the same time, these themes represent open opportunities for anyone who wants to exploit, contribute to, or expand on this research.
% \end{itemize}

\section{Related Work}
\label{sec:previous_work}
Automated game testing and validation has been gaining interest from both the research and video game development communities over recent years \cite{microsoft}. Here, we review work from the literature most related to our contribution.

\subsection{Automated Playtesting} 
Several studies have investigated the use of AI techniques for automated playtesting. Many of these works rely heavily on classical, hand-scripted AI or random exploration~\cite{pathos,personas}. The Reveal-More algorithm, in addition, uses human demonstrations to guide random exploration, although in simple 2D dungeon levels~\cite{reveal}. \citet{mugrai2019automated} developed an algorithm to mimic human behavior to achieve more meaningful gameplay testing results, but also to aid in the game design process.
Several approaches from the literature are based on model-based automated playtesting. \citet{platform} proposed a model-based testing approach for automated, black-box functional testing of platform games, and \citet{mobiletest} report their experience with developing model-based automated tests for 16 mobile games. Other notable examples of model-based testing can be found in~\cite{playfulness, crushinator, scenariobased, fifa}. However, when dealing with complex 3D environments these scripted or model-based techniques are not readily applicable due to the high-dimensional state space involved.

In parallel, many other works have used reinforcement learning (RL) to perform automated playtesting \cite{towardstest}. \citet{wuji} proposed an on-the-fly game testing framework which leverages evolutionary algorithms, deep-RL and multi-objective optimization to perform automated game testing. \citet{2dplaytest} trained reinforcement learning agents to perform automated playtesting in 2D side-scrolling games producing visualizations for level design analysis. \citet{augmenting} propose a study on the usefulness of using RL policies to playtest levels against model-based agents, while \citet{improving} used intrinsic motivation to train many agents to explore a 3D scenario with the aim of finding issues and oversights. Finally, \citet{ccpt} proposed the curiosity-conditioned proximal trajectories algorithm with which they test complex 3D game scenes with a combination of IL, RL and curiosity driven exploration. 

\subsection{Imitation Learning} Imitation learning (IL) aims to distill a policy mimicking the behavior of an expert demonstrator from a dataset of recorded demonstrations. It is often assumed that demonstrations come from an expert who is behaving near-optimally. Standard approaches are based on behavioral cloning that mainly use supervised learning~\cite{bc1,dagger,tamer}. Generative adversarial imitation learning (GAIL) is an imitation learning technique which is based on a generator-discriminator approach~\cite{gail}, as \citet{connection} noticed that imitation learning is closely related to the training of generative adversarial networks (GAN). Based on ideas from GAIL, \citet{amp} proposed the adversarial motion prior (AMP) algorithm, which is a GAN-based imitation learning method that aims to increase stability of adversarial approaches.

Few approaches have applied imitation learning to game testing agents. The method in the previously cited Reveal-More algorithm \cite{reveal} uses demonstrations to guide exploration, although it does not use a learning algorithm. \citet{winning} used an approach similar to behavioral cloning in which gameplay agents learn behavioral policies from the game designers. \citet{concurrent} trained agents through a combination of imitation learning and reinforcement learning with multi-action policies for a first person shooter game. \citet{adam} used an inverse reinforcement learning technique for training agents that can play a variety of games in the Atari 2600 game suite of the Arcade Learning Environment \cite{ale}. In the next section we describe why we think IL is a better approach to game validation than the cited approaches.

\section{Proposed Method}
\label{sec:method}

% \begin{table*}
% \centering
% \begin{tabular}{llll}
% \hline
% \textbf{Property}         & \textbf{Scripting}            & \textbf{RL}           & \textbf{IL} \\ \hline
% Setup time\footnote{Setup time: total time consumed setting up the behaviors. Training, demonstration time, scripting time, etc.). The values are taken from the examples given in section \ref{sec:exps}}              & Highly dependent*             & 1 h - 48 h\footnote{+ time for reward shaping (Alessandro: what did the experiment show training time was?)}              & 2 min - 5 min\footnote{Demonstration + training time}     \\
% Exploration             & no\footnote{Exploration has the be explicitly added so it will not create novel behaviors that are not already scripted} & yes & no\\ 
% Exploitation & no & yes & no \\
% Controllability & yes & no & yes \\
% Generalization & none & low & low \\
% ML knowledge required & no & yes & no \\
% Programming needed & yes & no & no \\
% \hline
% \end{tabular}
% \caption{Comparing parameters for scripting, RL and IL approaches.}
% \label{tab:comparing_approaches}
% \end{table*}

% The goal of this paper is to evaluate the validity of using a data-driven approach to behavior generation in order to aid and improve level design and gameplay validation. 
We propose an approach based on imitation learning to showcase the potential of data-driven methods for direct gameplay validation to designers. In this section, we first describe why we decide to use IL, then we detail our IL setup and training algorithm.

\subsection{Method Comparison}
% As previously mentioned, our hypothesis is that IL is a valid and efficient approach for game validation and level design. As cited in section \ref{sec:previous_work}, there are two main methods for automated playtesting currently used in both academia and industry: reinforcement learning and scripted model-based agents. In Table \ref{tab:comparing_approaches} we show a high-level summary of the properties of each respective approach.
Here we give a brief overview on the different common approaches to automated testing. 

\minisection{Reinforcement Learning.} RL for game validation is known to be sample-inefficient. For example, the work proposed by \citet{improving} required an average of two days of training before thorough coverage of a game scene was achieved. Even if the sample inefficiency can be mitigated by combining RL with IL similar to the curiosity-conditioned proximal trajectory algorithm \cite{ccpt}, it is still a challenge that must be addressed for RL to be practically useful in production. Moreover, RL typically leaves little control to the user over the final behavior of the policy as the trained agents will inherently exploit the reward function, regardless of the designers intention \cite{faultyreward}. We can alleviate this problem with reward shaping, but this requires considerable knowledge of RL and machine learning in general, making it unpractical to the general user. Even with these mentioned attributes, RL still has some advantages. If designers prefer exploitation and exploration over imitation then RL is preferable. And similar to IL, it does not need any programming knowledge to train. 
    
\minisection{Scripted Bots.} With scripted bots we mean bots modelled with programming methods, i.e. the standard way of creating game AI bots. For programming autonomous bots, designers need to have relatively high domain knowledge of the level design, but also scripting skills to successfully drive the behaviors of the bots. Moreover, even if they have very fine control over the bots final behavior, the scripted agents will quickly become sub-optimal, and even unusable, when they have to face design changes in the environment. The setup time also greatly depends on the complexity of the game and game scene. For the reason listed above, we argue that this approach does not satisfy the requirements listed in section \ref{sec:introduction}.

\minisection{Imitation learning.} IL is more sample-efficient than RL, and allows for training agents in minutes or only few hours \cite{ccpt}. With this method we only need to leverage the designer's domain knowledge of the game, thus effectively adding humans-in-the-loop for major controllability. Not only can we demonstrate the intended behavior, but also even correct it using new demonstrations with little additional training time. Compared to scripted behavior, IL provides the same level of generalization as RL as the policy model should be able to generalize in some extent to unseen observations. However, compared to both RL and model-based bots, IL calls for no or limited theory knowledge or programming skills as designers only have to demonstrate what they want the IL agent to do. With IL, we may mitigate some of the drawbacks mentioned of the two previous methods. 

\begin{figure}
    \begin{center}
    \includegraphics[width=0.8\columnwidth]{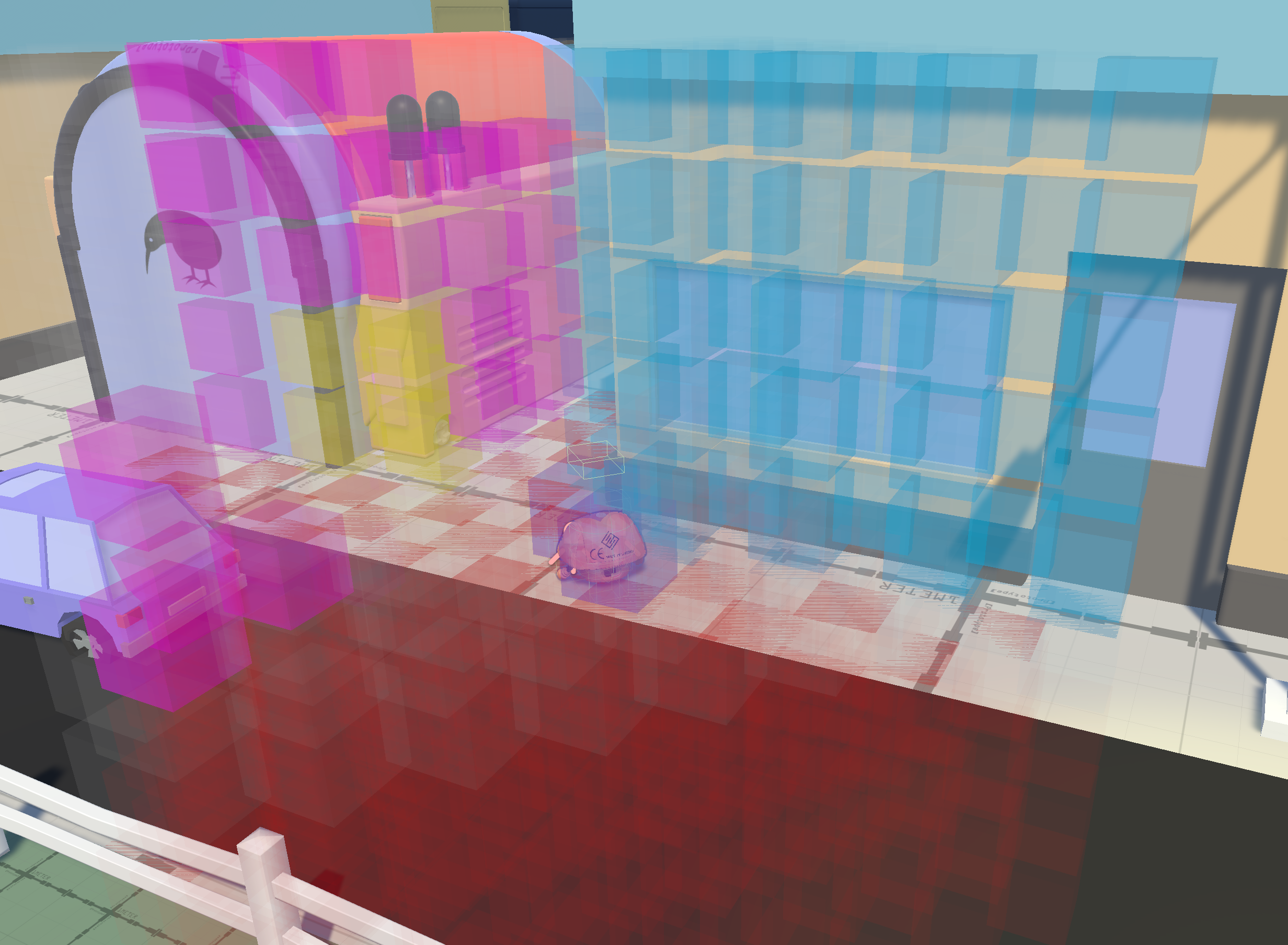}
    \end{center}
    \caption{Example of a semantic map used by the agent. Each cube describes the semantic integer value of the type of object at the corresponding position relative to the agent highlighted in blue.}
    \label{fig:semantic}
\end{figure}

\subsection{Approach}
Following the above argumentation, we developed a IL approach not only to demonstrate our claims, but also to showcase different use-cases for professional designers for helping  automated game validation.

\minisection{Algorithm.} We use an IL approach based on the DAgger algorithm \cite{dagger}, visible in Figure~\ref{fig:teasing}. DAgger allows designers to train agents interactively like they are actually playing the game. The core of the algorithm is to let designers seamlessly move between designing the level, providing game demonstrations and getting feedback from trained testing agents. Our approach allows real-time training as it requires low training time and it is more sample-efficient than the baseline methods, as we see in Section \ref{sec:exps}. Designers can also provide corrections to faulty agent behaviors, resulting in a continuous and instantaneous feedback loop between designers and agents. Providing corrections is as easy as taking the controller back and playing the game one more time. Once they are satisfied with the agents behavior, designers can stop providing demonstrations and watch the agent validate the game. Developers can then make any kind of design changes and wait for agent feedback, without making any new demonstrations.

\begin{figure}
    \begin{center}
    \includegraphics[width=0.9\columnwidth]{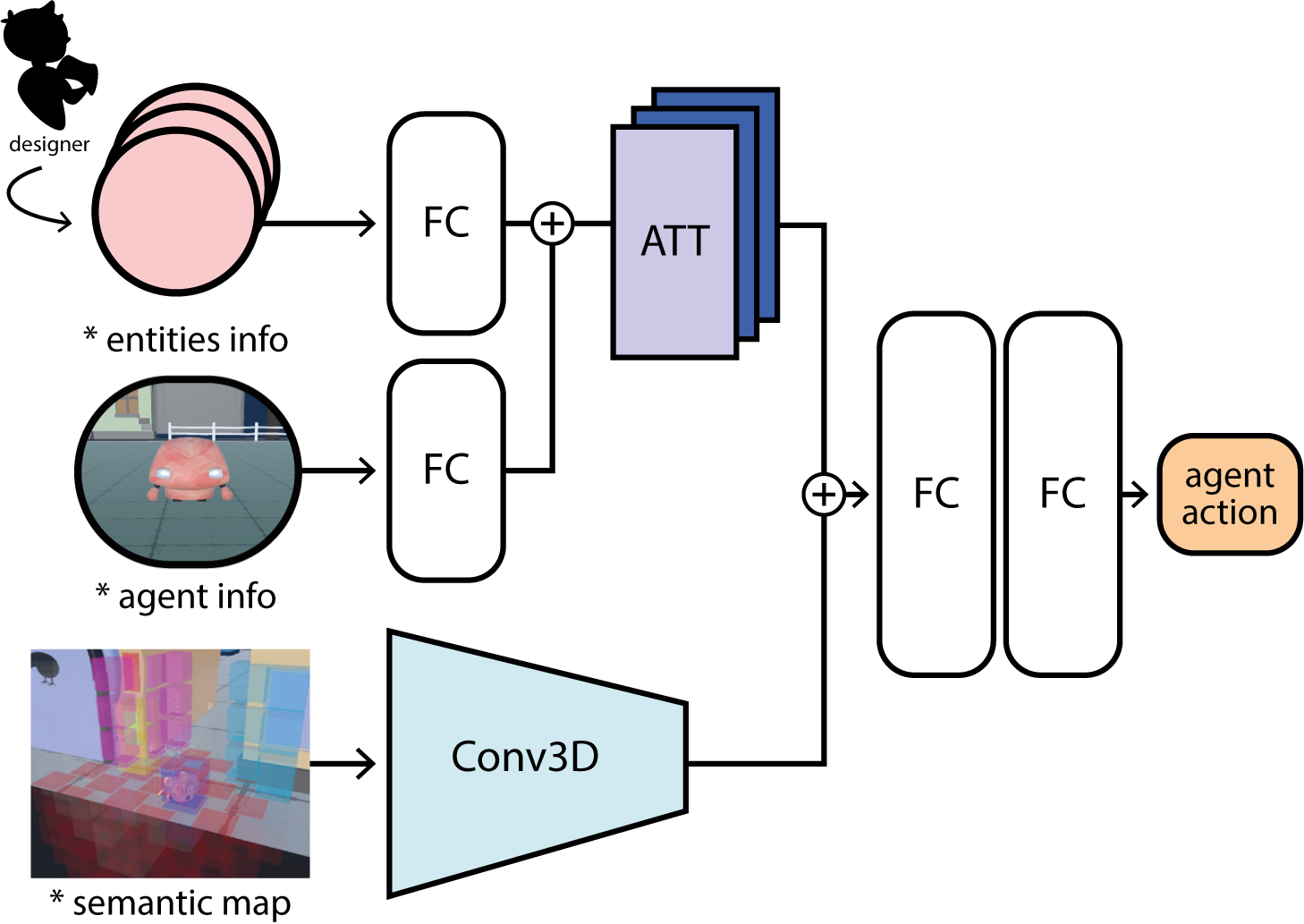}
    \end{center}
    \caption{Overview of the neural network architecture used in this work.}
    \label{fig:net}
\end{figure}

\minisection{State Space.} For the approach to be effective it needs to be as general as possible to adapt to many different game genres and scenarios that designers construct. Since 3D movements often are crucial gameplay elements of any video games, we focused on finding the best state representation taking this into consideration. This approach, with minor observation tweaking, will transfer well to similar, but contextually different game modes.

For setting up a behavior, developers define a goal position in the game environment. The spatial information of the agent relative to the goal is composed of the $\mathbb{R}^{2}$ projections of the agent-to-goal vector onto the $XY$ and $XZ$ planes as well as their corresponding lengths normalized to the gameplay area. We also include information about the agent indicating whether it can jump, whether it is grounded or climbing as well as any other auxiliary data which is expected to be relevant to gameplay. 
% For example, if the a game state like health of the agent should alter the behavior the model has to have this in its observation space. 
The user also specifies a list of entities and game objects that the agent should be aware of, e.g. intermediate goals, dynamic objects, enemies, and other assets that could be used for achieving the final goal. From these entities, the same relative information is inferred as for the main goal position relative to the agent. Lastly, the agent also has local perception. A semantic map solution is used similar to the one found by \citet{ccpt} which is general and performant. We illustrate an example of a semantic 3D occupancy map as input to the networks in Figure~\ref{fig:semantic}. This map is a categorical discretization of the space and elements around the agent, and each voxel in the map carries a semantic integer value describing the type of object at the corresponding game world position relative to the agent. The size, resolution and shape of the semantic map can be configured by the designers.

% We also provide an alternative state space, the impro one
\minisection{Neural Network.} We build the neural network $\pi_\theta$ with parameters $\theta$ with the goal of being as general and reusable as possible. First, all the information about the agent and goal is passed into a linear layer producing the self-embedding $x_\text{a} \in \mathbb{R}^{d}$, where $d$ is the embedding size. The list of entities is passed through a separate linear layer with shared weights producing embeddings $x_{\text{e}_i} \in \mathbb{R}^{d}$, one for each entity $\text{e}_i$ in the list. Each of these embedding vectors is concatenated with the self-embedding, producing $x_{\text{ae}_i} = [x_\text{a}, x_{\text{e}_i}]$, with $x_{\text{ae}_i} \in \mathbb{R}^{2d}$. This list of vectors is then passed through a transformer encoder with $4$ heads and final average pooling, producing a single vector $x_\text{t} \in \mathbb{R}^{2d}$. In parallel, the semantic occupancy map $M \in \mathbb{R}^{s \times s \times s}$ is first fed into an embedding layer, transforming categorical representations into continuous ones, and then into a 3D convolutional network. The output of this convolutional network is a vector embedding $x_\text{M} \in \mathbb{R}^{d}$ that is finally concatenated with $x_\text{t}$ and passed through an MLP, producing an action probability distribution. The complete neural network architecture is shown in Figure \ref{fig:net}.
Given a set of demonstrations:
$$D = \{\tau_i \; | \; \tau_i = (s_0^i, a_0^i, ..., s_T^i, a_T^i), \; i=1, .., N\},$$
we update the network following the objective:
\begin{equation}
\label{eq:bc}
    \arg\max_{\theta} \mathbb{E}_{(s,a) \sim \tau, \tau \sim D}[\log \pi_\theta(a | s)].
\end{equation}
This objective aims to mimic the expert behavior which is represented by the dataset $D$.

\begin{figure}
    \begin{center}
    \includegraphics[width=1.0\columnwidth]{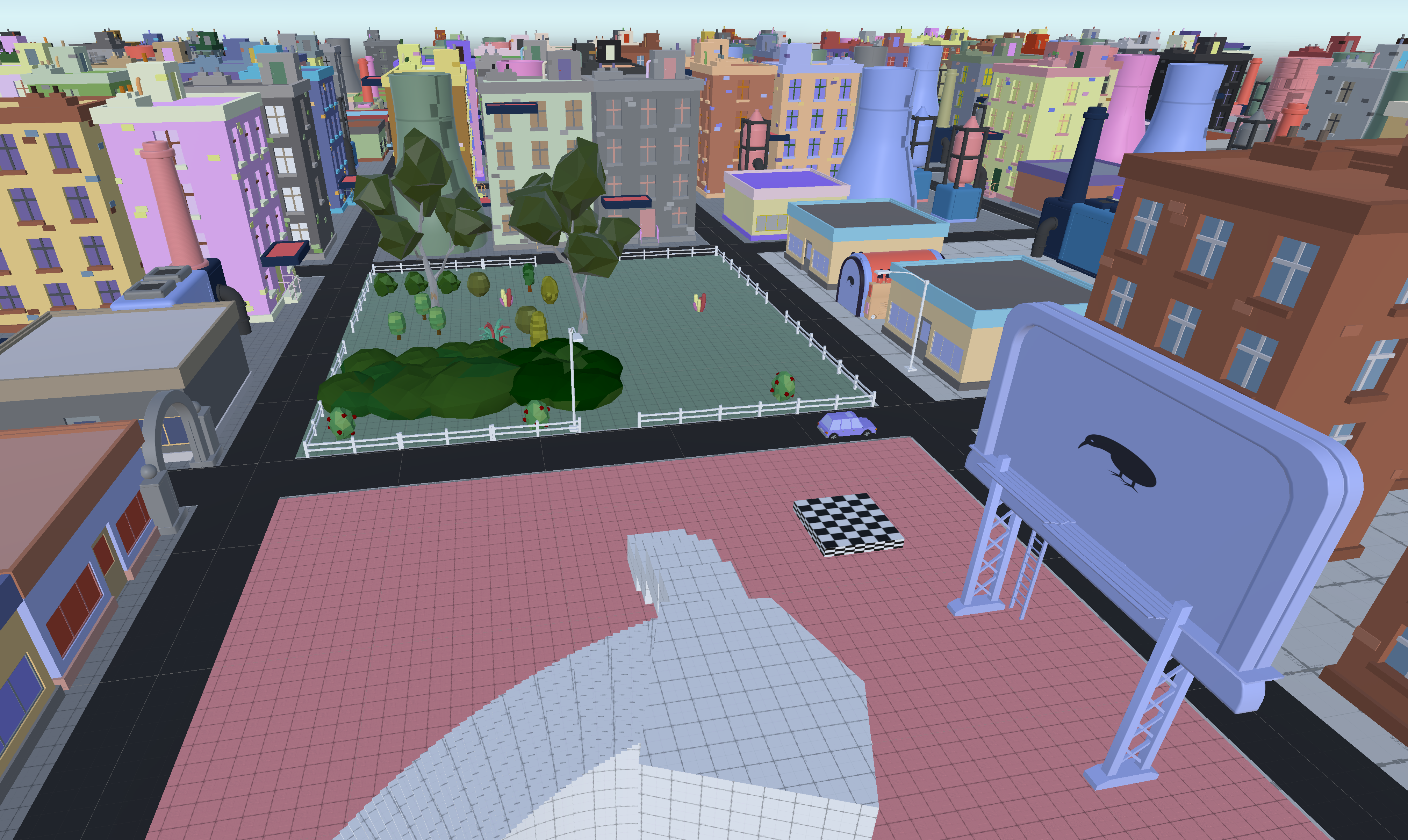}
    \end{center}
    \caption{Overview of the environment used in this study.}
    \label{fig:env}
\end{figure}

\section{Experiments}
\label{sec:exps}
In order to analyze the usefulness of our proposed approach, we conduct experiments demonstrating how IL can be utilized in different game design use-cases. First, we define the environment we use for our experiments, then we describe quantitative results of our approach compared to RL.

\subsection{Experimental Setup}
Here we describe the environment and training setup used in this work to test our proposed approach.

\minisection{Environment.}
The environment used in this work is shown in Figure \ref{fig:env}. This is a 3D navigation environment with procedurally generated elements. The environment is purposefully built like a mutable sandbox which can offer scenarios where a designer's level design workflow can be simulated. Users can add and change the goal location, agent spawn positions, layout of the level, location of intermediate goals and the locations of dynamic elements in the map such as elevators. In this environment, the agent has a set of $7$ discrete actions: move forward, move backward, turn right, turn left, jump, shoot and do nothing. In addition, the agent can use some interactable objects located around the map. 

\begin{figure}
    \begin{center}
    \includegraphics[width=0.9\columnwidth]{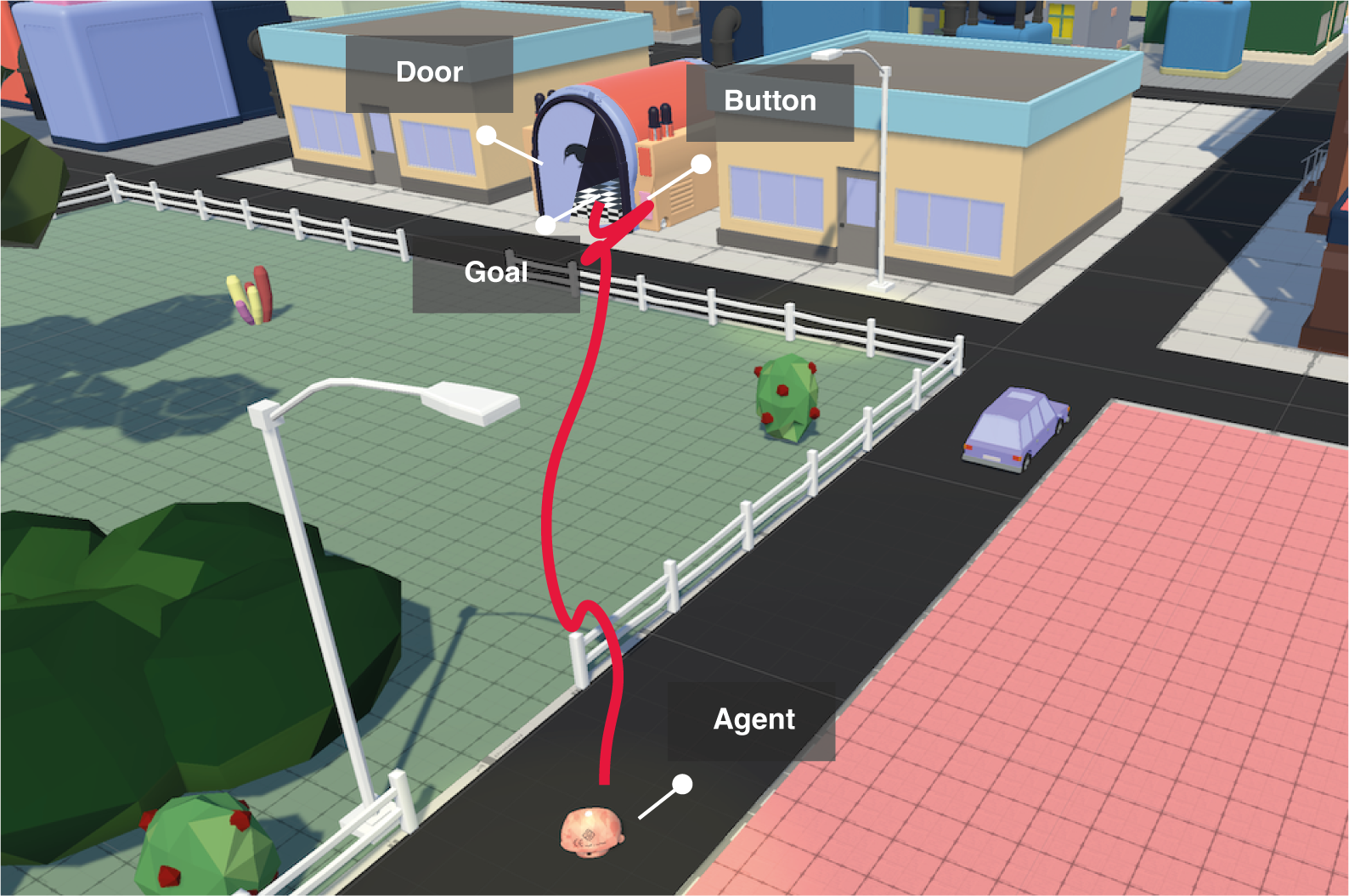}
    \end{center}
    \caption{Screenshot of the ``\textit{Validating Design Changes}'' use-case.}
    \label{fig:usecase1}
\end{figure}

\minisection{Training Setup.}
We compare our approach to two main baselines: Simple-RL and Tuned-RL, both using the PPO algorithm with identical hyperparameters \cite{ppo}. Simple-RL utilizes a naive reward function that gives a positive, progressive reward based on the distance to the goal. For Tuned-RL, a hand-crafted, dense reward function is used that is designed to lead the agent along a path, reaching multiple sub-goals before the final main goal. We use the same setting for all subsequent experiments. For each of the methods, we are interested in: the success rate of the agent (i.e. how many times it reaches the goal), training time (i.e. the time it takes to reach that success rate), generalization success rate (i.e. the success rate of the same agent in a different version of the environment), and imitation metric (i.e. how close the trajectories made by the agent are to the demonstrations). For the latter, we use the 3D Frechét distance between the agent trajectories and the demonstrator trajectories. All training was performed on a single machine with an NVIDIA GTX 1080 GPU with 8 GB VRAM, a 6-core Intel Xeon W-2133 CPU and 64 GB of RAM.

\subsection{Use-case Evaluation}
In order to evaluate the performance of the approach, we define a set of use-cases that resemble typical situations that game designers face in their daily workflow. In Table \ref{tab:quant_res}, we report quantitative results for each use-case, comparing our approach to the aforementioned RL baselines.

\minisection{Use-case 1: Validating Design Changes.} For the first use-case, we investigate if an IL agent in real-time can validate the layout of a level. The goal in this experiment is to train an agent to follow human demonstrations, then change the level and see how the agent adapts to these changes. This will give developers a glimpse into the usefulness of our proposed approach when they are still in the design process: they can see how players would adapt accordingly to modifications in the level layout.
In Figure \ref{fig:usecase1}, we show details of the use-case. The agent is tasked to navigate to a goal hidden behind a door that can be opened by interaction with a button. After training the agent, we test it by removing and adding new elements to the level. 

We argue, with the support of Table \ref{tab:quant_res}, that IL is more suitable than RL for use-cases like this. IL is generally significantly faster than RL, and at the same time it gives the user much more controllability over the final agent behavior as the agent can be guided via the demonstrations. As seen in this table, we could improve controllability and training time of the RL solution (e.g. Tuned-RL), but this requires reward shaping that is a very difficult task, especially for non-experts in machine learning. Moreover, defining a suitable reward function requires many adjustment iterations to the training setup that decreases the overall efficiency of the system. Furthermore, it is evident from the table that IL gives the same level of generalization as RL and it adapts easily to slight variations of the same environment.

\begin{figure}
    \begin{center}
    \includegraphics[width=0.9\columnwidth]{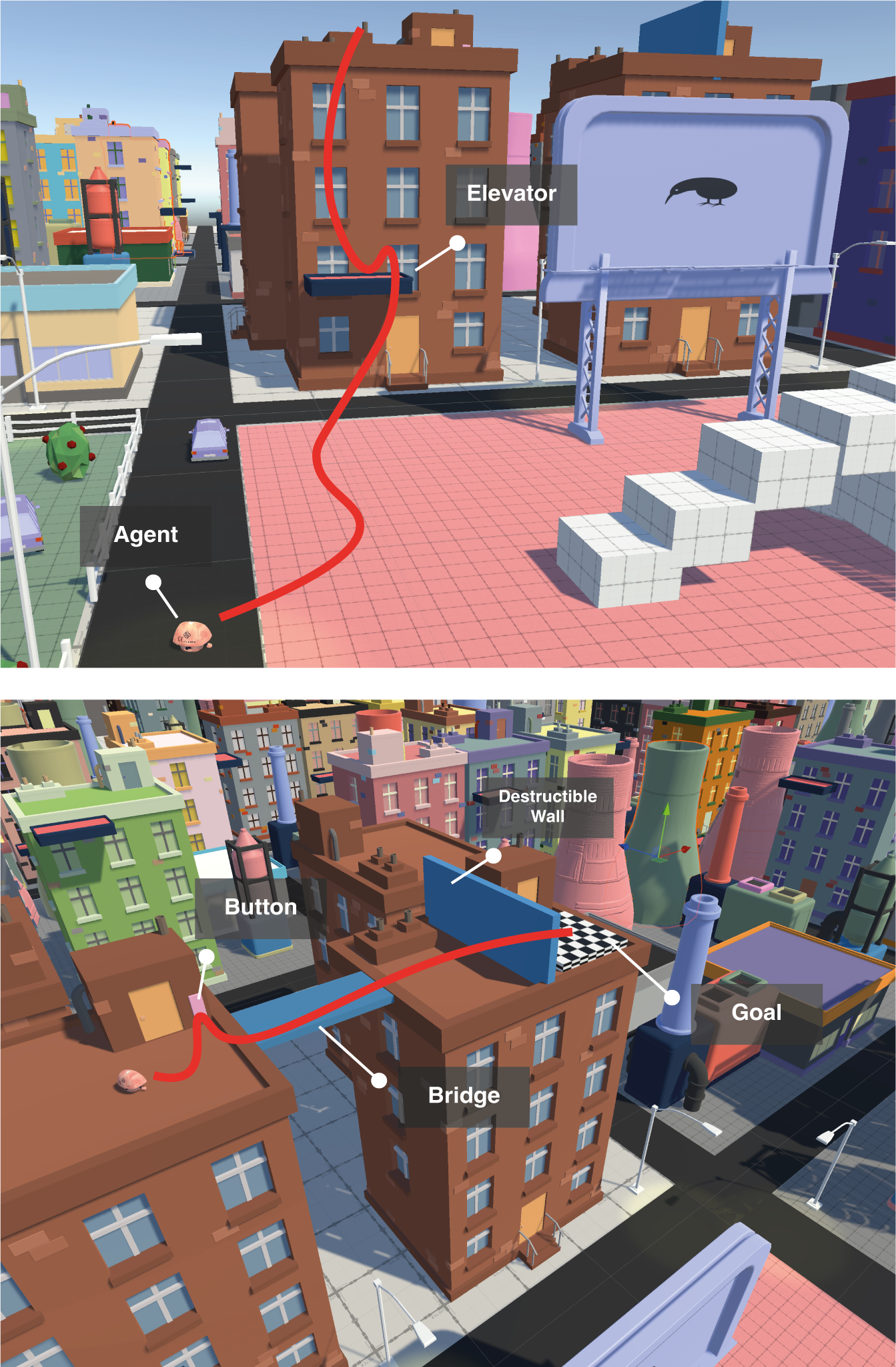}
    \end{center}
    \caption{Screenshot of the ``\textit{Validating a Complex Trajectory}'' use-case.}
    \label{fig:usecase2}
\end{figure}

\minisection{Use-case 2: Validating Complex Trajectories.} For this use-case, we train our agent to reproduce a complex trajectory shown in Figure \ref{fig:usecase2}. The agent has multiple intermediate goals: use the elevator, interact with the button to create the bridge, destroy a wall by shooting it and arrive at the goal location. Here we are not interested in generalization, but only on the ability to replicate the expert behavior as quickly as possible.

In Table \ref{tab:quant_res}, we can see how our interactive approach is much more suitable for our goals compared to the baselines. As the table shows, it takes a lot of time for the RL agent to be trained, which is problematic as efficiency is a key requirement. Even the Tuned-RL baseline is much time consuming than our approach. Moreover, the RL agent will exploit the environment without taking into account the real intention of the designer. An interesting finding here is that the RL agent has found a different way to get to the goal location, which is not desirable for this specific use-case but would be very useful for exploit detection.

\begin{figure}
    \begin{center}
    \includegraphics[width=0.95\columnwidth]{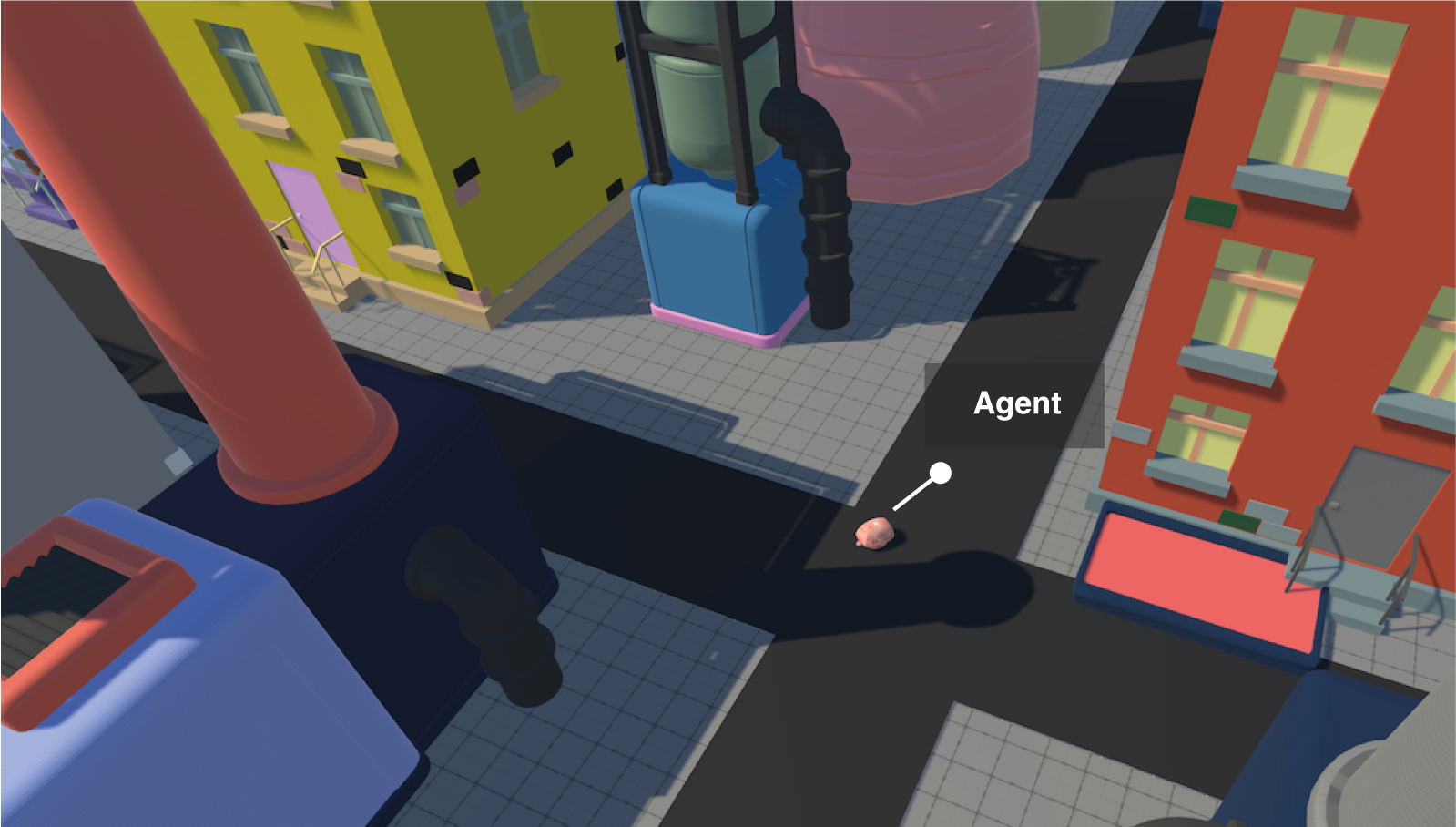}
    \end{center}
    \caption{Screenshot of the ``\textit{Navigation}'' use-case.}
    \label{fig:usecase3}
\end{figure}

\minisection{Use-case 3: Navigation.} As previously mentioned, navigation is one of the fundamental aspects in modern video games. However, the traditional scripted way to handle navigation is to pre-bake and use navigation meshes in combination with classical pathfinding algorithms. It is known that using a navigation mesh becomes intractable in many practical situations, and we thus must compromise between navigation cost and quality. This is accomplished in practice by either removing some of the navigation abilities or by pruning most of the navigation mesh connections \cite{ubisoftnavigation}. Moreover, every time designers change something in the layout, the navigation mesh needs to be regenerated. For this use-case, we consider a complex navigation task where the agent has to navigate a city that the designers have not completely finished yet. To simulate this, we first record demonstrations in an unfinished city, and then we test the trained agents within a test city that changes every 2 seconds until it reaches the goal location. A screenshot of this use-case is visible in Figure~\ref{fig:usecase3}. Similar to the second use-case, we want the agent to always follow the same path that is demonstrated.

Table \ref{tab:quant_res} summarizes the results. Since the city environment is large, training agents with RL would take a lot of time compared to guiding them  along the correct path with our approach. Moreover, even if they enjoy a similar level of generalization, the corresponding RL agents will try to exploit and explore the environment finding different ways to arrive to the goal location. For our purposes we not only need generalization, but also a certain level of imitation, controllability and efficiency. The Tuned-RL baseline, however, achieves slightly better results in this case. However, we must reiterate that this baseline is infeasible to replicate by a game designer as it needs a very specific reward function and many iterations. Even then, the overall result is not very different from our approach. As shown by the results from these experiments, the combination of these elements is more easily achievable with IL rather than RL. 

%%% Ablation studies for the architecture: w/ and w/o transformers

With these analyses, we wanted to assess our hypothesis detailed in Section \ref{sec:method} that data-driven modelling via IL is the preferred solution for achieving interactive game design and validation compared to the baselines. Next, we evaluate our proposed approach using feedback from professional game designers.

% \renewcommand{\tabcolsep}{2pt}
% \begin{table*}
% \centering
% \begin{tabular}{l|llll|llll|llll}
% \hline
%  & \multicolumn{4}{c|}{Use-case 1} & \multicolumn{4}{c|}{Use-case 2} & \multicolumn{4}{c}{Use-case 3} \\ 
%  &Success $\uparrow&Success& Time $\downarrow$ & Generalization $\uparrow$ & Imitation $\downarrow$  &Success $\uparrow&Success& Time $\downarrow$ & Generalization $\uparrow$ & Imitation $\downarrow$  &Success $\uparrow&Success& Time $\downarrow$ & Generalization $\uparrow$ & Imitation $\downarrow$  \\
% \hline
% \textbf{Ours} & {$\mathbf{0.95 \pm 0.00}$} & $\mathbf{0.02} \text{ h}$ & {$\mathbf{0.96 \pm 0.05}$} & $\mathbf{6.84 \pm 0.34}$ & $0.90 \pm 0.02$ & $\mathbf{0.06} \text{ h}$ & - & $\mathbf{7.73 \pm 1.10}$ & $0.81 \pm 0.08$ & $\mathbf{0.22  \text{ h}}$ & $0.78 \pm 0.03$ & $46.07 \pm 1.03$ \\ 
% RL & $0.91 \pm 0.02$ & $5.00 \text{ h}$ & $0.90 \pm 0.00$ & $8.37 \pm 0.27$ & $0.00 \pm 0.00$ & $18.18 \text{ h}$ & - & $28.11 \pm 0.41$ & $0.00 \pm 0.00$ & $16.49 \text{ h}$ & $0.00 \pm 0.00$ & $80.22 \pm 0.53$ \\
% SRL & $0.95 \pm 0.00$ & $4.48  \text{ h}$& $0.90 \pm 0.02 $ & $8.13 \pm 0.20$ & $\mathbf{0.92 \pm 0.03}$ & $13.56 \text{ h}$& - & $9.53 \pm 1.37$ & $\mathbf{0.86 \pm 0.05}$ & $4.12 \text{ h}$ & $\mathbf{0.87 \pm 0.03}$ & $\mathbf{16.79 \pm 2.12}$ \\
% \end{tabular}
% \caption{Quantitative results of our experiments. }
% \label{tab:quant_res}
% \end{table*}

\renewcommand{\tabcolsep}{3pt}
\begin{table}
\centering
\begin{tabular}{l|llll}
\hline
& \multicolumn{4}{c}{Use-case 1} \\
&Success $\uparrow$& Time $\downarrow$ & Generalization $\uparrow$ & Imitation $\downarrow$  \\
\hline
\textbf{Ours} & {$\mathbf{0.95 \pm 0.00}$} & $\mathbf{0.02} \text{ h}$ & {$\mathbf{0.96 \pm 0.05}$} & $\mathbf{6.84 \pm 0.34}$ \\
Simple-RL & $0.91 \pm 0.02$ & $5.00 \text{ h}$ & $0.90 \pm 0.00$ & $8.37 \pm 0.27$ \\
Tuned-RL & $0.95 \pm 0.00$ & $4.48  \text{ h}$& $0.90 \pm 0.02 $ & $8.13 \pm 0.20$ \\
\hline
& \multicolumn{4}{c}{Use-case 2} \\
&Success $\uparrow$& Time $\downarrow$ & Generalization $\uparrow$ & Imitation $\downarrow$  \\
\hline
\textbf{Ours} & $0.90 \pm 0.02$ & $\mathbf{0.06} \text{ h}$ & - & $\mathbf{7.73 \pm 1.10}$ \\
Simple-RL & $0.00 \pm 0.00$ & $18.18 \text{ h}$ & - & $28.11 \pm 0.41$ \\
Tuned-RL & $\mathbf{0.92 \pm 0.03}$ & $13.56 \text{ h}$& - & $9.53 \pm 1.37$ \\
\hline
& \multicolumn{4}{c}{Use-case 3} \\
&Success $\uparrow$& Time $\downarrow$ & Generalization $\uparrow$ & Imitation $\downarrow$  \\
\hline
\textbf{Ours} & $0.81 \pm 0.08$ & $\mathbf{0.22  \text{ h}}$ & $0.78 \pm 0.03$ & $46.07 \pm 1.03$ \\ 
Simple-RL & $0.00 \pm 0.00$ & $16.49 \text{ h}$ & $0.00 \pm 0.00$ & $80.22 \pm 0.53$ \\
Tuned-RL & $\mathbf{0.86 \pm 0.05}$ & $4.12 \text{ h}$ & $\mathbf{0.87 \pm 0.03}$ & $\mathbf{16.79 \pm 2.12}$ \\
\hline
\end{tabular}
\caption{Quantitative results of our experiments. We compare our approach with two main baselines: Simple-RL, which uses PPO algorithm \cite{ppo} with a sparse reward function and Tuned-RL which uses an hand-crafted and very dense reward function. All the numbers refer to the mean and standard deviation of $5$ training runs. For a complete description of the use-cases, see Section \ref{sec:exps}. Since in use-case 2 we are not interested in generalization, we do not report the values for that experiment.}
\label{tab:quant_res}
\end{table}

\section{User Study}
\label{sec:qual_res}
We performed a user study in the form of an online survey with professional game and level designers not only to assess desirability of using our proposed approach, but also to identify open opportunities for supporting automated game validation.

\subsection{Survey Description}
\label{sec:survey}
Our methodology consists of qualitative data collection applying an online survey
% which questions are publicly available\footnote{Link to survey questions}. 
The data for the survey were collected from professional game and level designers, recruited using snowball sampling. Subjects were recruited among different game studios of varying sizes, with different background knowledge and workflows. Table \ref{tab:participants} summarizes participant details. 
The survey is composed of Likert questions with some additional open questions. We also let participants add whatever feedback they want to individual questions.

\begin{table}
\centering
\begin{tabular}{lllll}
\hline
\textbf{ID} & \textbf{Role} & \textbf{Genre(s)} & \textbf{YoE} & \textbf{MLK}\\ \hline
P01 & Level Designer & FPS & 15 & Low \\ %peter.vestifrendrup
P02 & Level Designer & FPS & 11 & None \\ %fredrik.englund 
P03 & Level Designer & Racing & 3 & Very Low \\ %peter.vestifrendrup
P04 & Level Designer & RPG & 6 & High \\ %jponzio
P05 & Level Designer & Racing & 1 & Very Low \\ %aidan.coxon
P06 & Level Designer & RPG & 4 & Very Low \\ %rmann
P07 & Game Designer & RPG & 9 & Low \\ %jstramaglia
P08 & Game Designer & Sport & 3 & Very Low \\ %ygomes
P09 & Game Designer & Sport & 4 & Very Low \\ %mkrulevich
P10 & Game Designer & Match3 & 1 & Very Low \\ % ttolstoy
P11 & Level Designer & RPG & 15 & Very Low \\ %etaylor
P12 & Level Designer & FPS & 21 & Low \\ %ehanes
P13 & Gameplay Designer & RPG & 15 & Very Low \\ %ddemaree
P14 & Game Designer & RPG & 22 & None \\ %preston
P15 & Level Designer & FPS, racing & 10 & Very Low \\ % jonathan
P16 & Level Designer & RPG & 5 & Very Low \\ %  mhepguler
% P16 & Game AI Engineer & FPS, TPS & 12 & High \\ %jonas
\hline
\end{tabular}
\caption{Summary of participants. Abbreviations: YoE (years of experience), MLK (machine learning knowledge), FPS (first person shooter), RPG (role-playing game), TPS (third person shooter).}
\label{tab:participants}
\end{table}

The survey was divided into four sections: the first asks participants of some background information; the second asks them about their current game validation workflow, in particular if they use manual or automated playtesting; the third is the main part of the survey, as it asks participants what they think about our solution, if they would use it in their games, what characteristics an agent/approach like this should have to help them in their game and level design work, if they think IL would help them create better games; and the fourth contains optional questions about possible use-cases and future directions they see that we had not considered.

Since one of the main focuses of this paper is to assess what improvements and research are needed to maximize the value of using such an approach, one important question in section 3 of the survey asks participants to evaluate various characteristics that an agent for automated content evaluation should have. These characteristics are:
\begin{itemize}
    \item \textit{Imitation}: the agent can exactly replicate the demonstrations;
    \item \textit{Generalization}: the agent can adapt to different variations of the same situation;
    \item \textit{Exploration}: the agent can explore beyond the demonstrations and find bugs and issues;
    \item \textit{Personas}: the agent can have different types of behaviors;
    \item \textit{Efficiency}: the agent must use as few demonstrations as possible;
    \item \textit{Controllability}: having control over the agent behavior vs complete autonomy;
    \item \textit{Feedback}: the agent gives feedback when the amount of demonstration data is enough;
    \item \textit{Fine tuning}: behaviors can be fine tuned after initial training;
    \item \textit{Interpretability}: the agent can inform me when and why it fails.
\end{itemize}

\subsection{Survey Results}
\label{sec:survey_res}
We had a total of 16 accepted responses. The majority (71.4\%) of the participants have more than 5 years of experience in level design, with a median of 7.5 years. All the participants were level, game or gameplay designers. We found that the general machine learning knowledge of the participants is from very low to low. 83.3\% of the respondents do not use automatic validation of their levels and they rely on external people or systems for their level validation. 72.2\% of designers never rely on automated rather than manual playtesting and only 38.9\% have used at least one a scripted automated method in their daily work. They work on different game projects and game genres and they use different tools in their level design workflow, but all respondents have knowledge of and use game engine editors. 

\begin{figure}
    \begin{center}
    \includegraphics[width=1\columnwidth]{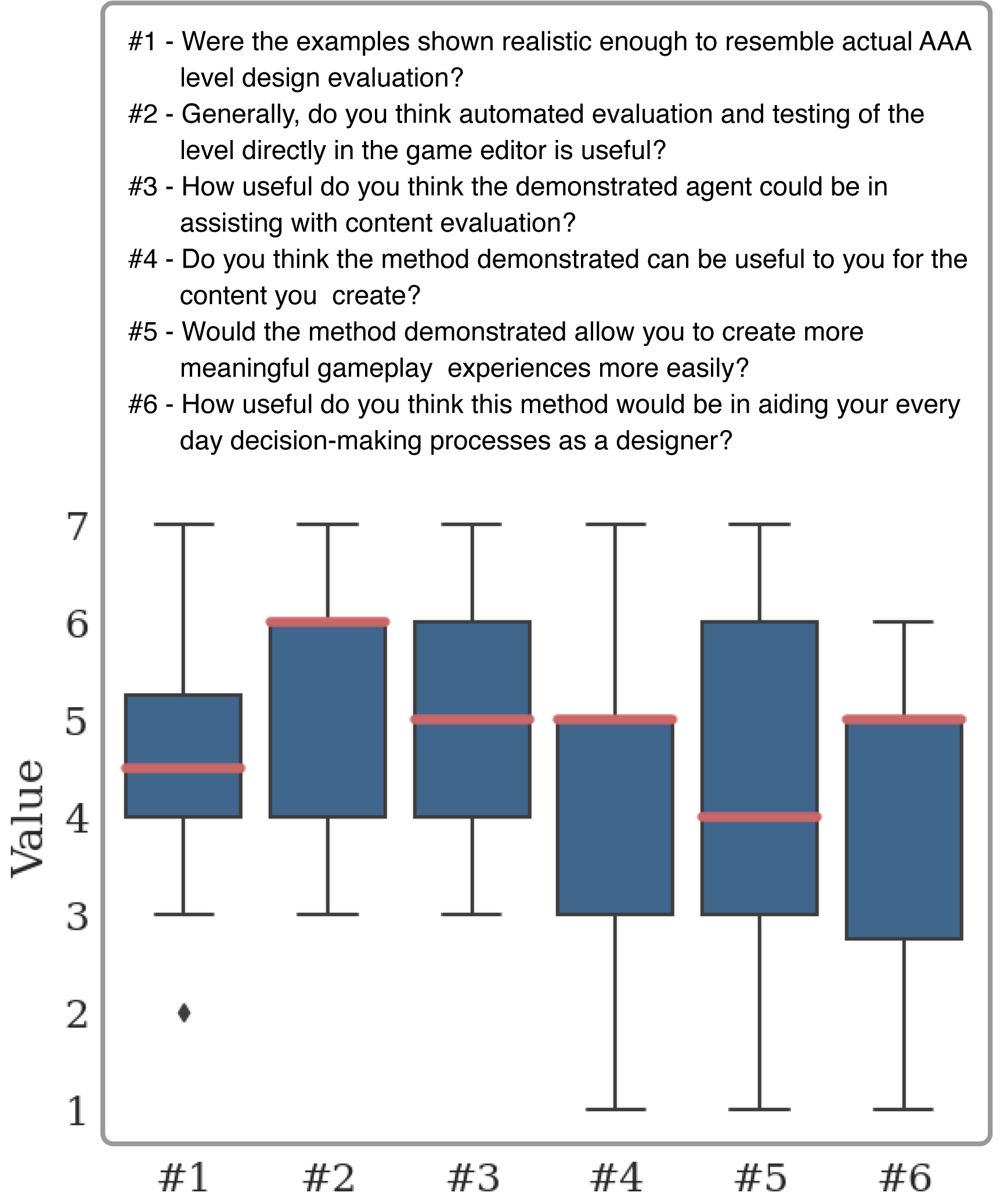}
    \end{center}
    \caption{Boxplot of some of the most important questions of the survey. The boxes represent the area between the lower quartile and the upper quartile. The red lines represent the median of the values, while the whiskers represent the minimum and maximum of the distributions, except for the outliers that are visualized with black dots.}
    \label{fig:qual_res}
\end{figure}

In Figure \ref{fig:qual_res} are shown some of the most interesting results we got from the survey. The general feeling is that designers would very likely use a tool like this as \textit{``it definitely would be useful to speed up the typically time consuming iterative design process''} and \textit{``a method like this can definitely speed up iteration, which is one of the main things any designer spends a lot of time''}. Moreover, they can relate to the demonstrated example as ``\textit{it is not quite like how they are building their levels, but it is not far off}'' and ``\textit{they think [the examples shown are] realistic for some games such as platformers}''. 

Participants acknowledge the usefulness and the potential of the demonstrated method, even if it will probably need the proper adaptations for the specific game genre they are working on. The 81.2\% of the respondents agree that automated validation of the level directly in the game editor is useful and 93.0\% of them think the demonstrated agents could be useful for assisting in their content validation. Many respondents (41.1\%) agree that a method like this adds value compared to scripted agents and some of them (23.5\%) think it could completely replace scripting. 

Some designers were skeptical about the complexity of the examples shown by the survey and they would like to see what happens in more complicated game mechanics. One participant says that \textit{``[he is] wondering about how effective this would be in genres that are more complicated mechanically''}. This was expected due to the preliminary nature of this study and that we do not cover every genre but focus on certain gameplay. However, a larger group (81.2\%) though that complexity was high enough which reinforces the notion that the environment is relevant to a significant part of the community. Designers also describe the challenges of using such a tool in their daily workflow: one argue that ``\textit{human players play very differently then bots, can be difficult to only rely on AI when it comes to testing}'' or ``\textit{demonstrations take time, which would push busy level designers to not use it. Demonstrating how to solve specific issues (how to go around a car, or recover from a mistake) should not be on the level designer}''. Not all game creators were fully convinced by the approach because ``\textit{the demonstration video showed a process that does not really show a level designer anything new or unexpected}''.

After describing their doubts, we asked participants to evaluate each of the characteristics delineated in the previous section and which are the most important ones to improve the approach. Table \ref{tab:char} refers to the results. We claim that these results bring much value to our research: not only do we know that our initial hypothesis is supported by professional designers, but we also know what they really want and why they are skeptical of using such an approach. This is important because, as we will see in Section \ref{sec:future}, the values in Table \ref{tab:char} form very precise research directions that we encourage all of the game research community to consider in future contributions to the video game industry.

\section{Future Work}
\label{sec:future}
Results in section \ref{sec:survey_res}, and in particular Table \ref{tab:char}, showcased challenges of using IL as a game validation tool. Here we propose research opportunities and future work that we gained from interacting and interviewing experts in the field. With this we want to lower the bar for researcher who would like to contribute and improve said approach, as this would help drive the research, and therefore industry, forward.

\minisection{Generalization.} From the survey, it is clear that one of the most important request is to have an agent that not only imitates the expert behavior, but that is more representative of the unpredictable nature of the players. Moreover, ``\textit{demonstrating how to solve specific issues should not be on the level designer}''. Humans can easily understand, given a previous task demonstration, how to adapt to slight changes in the environment or to recover from faulty situations which in comparison is difficult for machine learning models. With our proposed general purpose neural network and state space we mitigate this problem, but in many cases agents trained under one set of demonstrations are not able to adapt to slightly different situations \cite{adversarialattack}. Generalization as a subject in self-learning agents has already been addressed in the literature, but the state-of-the-art approaches are not readily applicable for our purpose: most of them use either interactions with the environment \cite{gail, airl, bco} or learn the inverse dynamic of the agent \cite{abco}. Moreover, a recent study by \citet{ongenadv} proves that adversarial imitation learning algorithms do not generalize more than standard behavioral cloning in some domains. A possible future direction is to try data augmentation techniques used in offline reinforcement learning algorithms such as the work done by \cite{s4rl}. 
Offline reinforcement learning has several connections to behavioral cloning \cite{minimalist}. It learns a policy using a pre-defined dataset without direct interactions with the environment, but in contrast with imitation learning that supposes the data comes from an optimal policy, offline reinforcement learning relies also to sub-optimal examples. This lets trained agents to be more general and allows a higher exploration level. 
With all this considered, we argue that generalization is far from solved for this use case. Therefore, we conclude that research in this direction is one of the more effective way to improve the results of this approach.

\minisection{Personas.}  One feature requested recurrently by survey participants is the possibility to train the model for different behavior types, or so called personas. The aim is to have various behaviors similar to the multi-modal nature of how humans play, in order to create more meaningful agents. Research community has addressed the problem of creating different personas many times, but exclusively in RL contexts. Works from \citet{ubisoft_conditioned}, \citet{directrew} and \citet{policyfusion} have studied and explored how to learn and combine different behavior styles with different reward functions. This makes it challenging to train different styles with only IL. The work by \citet{amp} is an example on how to combine IL and RL to create different animation styles. More research into training different playstyles with IL only would allow designers to train more meaningful testing agents to validate different gameplay.

\minisection{Exploration.} At the same time participants frequently brought up the exploration aspect, i.e. the ability for the agent to look beyond expert demonstrations in search for bugs and overlooked issues. Exploration is a well known problem in RL literature \cite{rnd}, but exploration in IL is, to our knowledge, to a large extent a unexplored field. This is mainly due to the conflict nature of an agent that must both learn to follow more or less precisely the expert demonstrations as well as explore beyond the optimal behavior. Moreover, we would not use exploration to improve agents, but we would like to exploit exploration to find bugs. \citet{ccpt} provide a good example on how to leverage both IL and RL to train agents to both follow demonstrated behaviors but also to explore in search for overlooked issues. However, this type of solution is still sample inefficient to be used as an active game design tool. Research in this direction would likely help the utility of such tool for game design validation.

% \minisection{Personas and Exploration.} One feature requested recurrently by survey participants is the possibility to train the model for different behavior types, or so called personas. One is exploration, i.e. the ability for the agent to look beyond expert demonstrations in search for bugs and overlooked issues. The other one is having different behaviors similar to the multi-modal nature of how humans play, in order to create more meaningful agents. Exploration is a well known problem in reinforcement learning literature \cite{rnd}, but exploration in IL is, to our knowledge, a to a large extent a unexplored field. Moreover, we would not use exploration to improve agents, but we would like to exploit exploration to find bugs. \cite{ccpt} provide a good example on how to leverage both IL and RL to train agents to both follow demonstrated behaviors but also to explore in search for overlooked issues. However, this type of solution is still sample inefficient to be used as an active game design tool. At the same time, many have explored how to learn and combine different behavior styles \cite{ubisoft_conditioned, policyfusion, directrew}, however, it is unclear how to use these techniques with IL algorithms such as behavioral cloning and DAgger.

\minisection{Usability.} The usability aspect is one of the most important for participants. This includes both providing useful information to designers about the models they are training, but also the ease-of-use of the tool. Recent techniques from the explainable AI \cite{xairl} research community applied to RL could be used to address the challenge of interpreting and explaining behaviors of the agent. One can use techniques from game analytics research and gameplay visualization techniques \cite{vissurvey, ccpt}. At the same time, a sentiment found in the survey is that to be a usable tool there is a need to minimize the effort of the end-user. This includes getting feedback (e.g. demonstration samples needed, when to stop producing demonstrations), and more direct improvements to use as few demonstration samples as possible. 
To address the latter, few- and one-shot imitation learning is a recent active topic that potentially can help to reduce samples required. Works from \citet{oneshotil} and \citet{fewshotil} specifically address the problem within IL, mainly exploiting meta-learning techniques \cite{maml}. However, these state-of-the-art approaches require many preliminary training iterations and it is unclear how to make this in a game that still is not readily stable and not yet finished. One way to both improve agents quality and the usability of the tool is to leverage the already cited offline reinforcement learning techniques \cite{conservative, decision, s4rl}. We can leverage the recorder to not only store demonstrations, but also all the other interactions that the agent does while testing the level. In this way we can both improve the agents, be more sample efficient and improve generalization without the need to directly use the environment.
% Beyond this there are several other ways to improve usability of this approach and this only covers a few. We hope this anyway can spark some ideas on where there are room for improvements for this approach that is not solved in this paper.

\minisection{Multi-Agent.} Many designers noted that most of modern video games are multi-agent systems, and in order to thoroughly test these environments we need to train multiple interacting agents. However, most of the current research in IL focuses on single agents learning from a single teacher. Few examples of multi-agent imitation learning are: \citet{concurrent} use IL in a multi-agent game, but they do not really address the problem of a multi-agent system, while \citet{coordmail} propose a joint approach that simultaneously learns a latent coordination model along with the individual policies. We believe there is still a long way to go for multi-agent imitation learning, especially for this use-case, and we encourage researchers to follow this research direction.

\begin{table}
\centering
\begin{tabular}{llll}
\hline
 & \textbf{Mean} & \textbf{Median} & \textbf{Std} \\ \hline
Imitation & 4.56 & 4.00 & 1.63 \\
Generalization & 6.50 & 7.00 & 0.73 \\
Exploration & 6.81 & 7.00 & 0.40 \\
Personas & 5.87 & 6.00 & 1.31 \\
Efficiency & 5.50 & 5.50 & 1.21 \\
Controllability & 4.18 & 4.00 & 1.68 \\
Feedback & 6.00 & 6.00 & 0.96 \\
Fine tuning & 5.62 & 6.00 & 1.50 \\
Interpretability & 6.50 & 7.00 & 1.03 \\
\hline
\end{tabular}
\caption{Results of characteristics questions. Participants could give a value between $[1, 7]$ to each of the category, where $1$ means ``\textit{not so important}'' and $7$ means ``\textit{very important}''.}
\label{tab:char}
\end{table}

\section{Conclusion}
The use of machine learning has recently gaining attention in game industry for creating better quality games. However, many works for creating autonomous agents from the literature do not have an easily interpretable translation into an actual design tool as they need to satisfy certain requirements that usually the research community tends not to address. 
% Therefore, mass adoption of such tools by game developers requires significant technical innovation to satisfy these requirements.

In this paper, we first claimed that data-driven programming via imitation learning is a suitable approach for real-time validation of game and level design. We proposed an imitation learning approach and we investigated it focusing on three different design validation use-cases. We demonstrated how this type of approach can satisfy many of the requirements for being an effective game design tool in comparison to simple reinforcement learning and model-based scripted behaviors. We also performed a user study with professional game and level designers from different, and in many cases, disparate game studios and game genres. We asked participants to assess the desirability and opportunities of using such an approach in their daily workflow. Moreover, we asked designers what characteristics they would want from a data-driven tool for creating autonomous agents that validate their design.

The user study showcased the desire of designers to have an automated way to test and validate their games. Along with our preliminary results we proved that the data-driven approach we propose is a potential candidate for achieving such objectives. However, the study also highlighted challenges and the gap that exists between techniques from the literature and their actual use in the game industry. For this reason, we proposed a series of research directions that will help such an approach to move from an impractical tool to an effective game design tool. We expect that our recommendations will foster game research community to contribute to or expand on this research.

\bibliographystyle{IEEEtranN}
{\footnotesize \bibliography{refs}}

\end{document}